\documentclass[fleqn,usenatbib]{mnras}

\usepackage[T1]{fontenc}
\usepackage{lscape}

\DeclareRobustCommand{\VAN}[3]{#2}
\let\VANthebibliography\thebibliography
\def\thebibliography{\DeclareRobustCommand{\VAN}[3]{##3}\VANthebibliography}

\usepackage{graphicx}	
\usepackage{amsmath}	
\usepackage{amssymb}	
\usepackage[normalem]{ulem}
\usepackage{tablefootnote}

\title[Damping of slow waves along umbral fan loops]{Effect of area divergence and frequency on damping of slow magnetoacoustic waves propagating along umbral fan loops}

\author[Rawat \& Gupta]{Ananya Rawat$^{1,2}$\thanks{E-mail: ananyarawat@prl.res.in (AR)}, Girjesh R. Gupta$^{1}$\thanks{E-mail: girjesh@prl.res.in (GRG)}
\\
$^{1}$ Udaipur Solar Observatory, Physical Research Laboratory, Dewali, Badi Road, Udaipur 313001, India \\
$^{2}$ Department of Physics, Indian Institute of Technology Gandhinagar, Palaj, Gandhinagar 382355, India \\
}

\date{Accepted . Received ; in original form}

\pubyear{2023}

\begin{document}
\label{firstpage}
\pagerange{\pageref{firstpage}--\pageref{lastpage}}
\maketitle

\begin{abstract}
Waves play an important role in the heating of solar atmosphere, however, observations of wave propagation and damping from the solar photosphere to corona through chromosphere and transition region are very rare. Recent observations have shown propagation of 3-min slow magnetoacoustic waves (SMAWs) along fan loops from the solar photosphere to corona. In this work, we investigate the role of area divergence and frequencies on the damping of SMAWs propagating from the photosphere to the corona along several fan loops rooted in the sunspot umbra. We study the Fourier power spectra of oscillations along fan loops at each atmospheric height which showed significant enhancements in 1--2 min, 2.3--3.6 min and 4.2--6 min period bands. Amplitude of intensity oscillations in different period bands and heights are extracted after normalizing the filtered light curves with low-frequency background. We find damping of SMAW energy flux propagating along the fan loop 6 with damping lengths $\approx170$ km and $\approx208$ km for 1.5-min and 3-min period bands. We also show the decay of total wave energy content with height after incorporating area divergence effect, and present actual damping of SMAWs from photosphere to corona. Actual damping lengths in this case increases to $\approx172$ km and $\approx303$ km for 1.5-min and 3-min period bands. All the fan loops show such increase in actual damping lengths, and thus highlight the importance of area divergence effect. Results also show some frequency-dependent damping of SMAW energy fluxes with height where high-frequency waves are damped faster than low-frequency waves. 
\end{abstract}



\section{Introduction}
\label{sec:intro} 

Chromospheric and coronal energy losses in the solar active regions are approximately $2\times10^7$ and $10^7$ erg cm$^{-2}$ s$^{-1}$, respectively \citep{1977ARA&A..15..363W}. To maintain the hot atmosphere of the Sun, magnetohydrodynamic (MHD) waves is one of the proposed mechanisms to transfer energy into the upper solar atmosphere \citep[see recent review by][]{2020SSRv..216..140V}. These waves are excited by the interaction of magnetic field lines and convection motions at the photosphere. The available energy flux provided by the photosphere is about $1.2\times10^9$ erg cm$^{-2}$ s$^{-1}$ for a mean field strength of 500 G \citep{2012RSPTA.370.3217P}. The transfer and conversion of this mechanical energy flux into heat in the upper atmosphere \citep[e.g.,][]{2007ApJS..171..520C,2011ApJ...736....3V} is still not completely understood and is commonly called the coronal heating problem. Henceforth, the photosphere can provide sufficient energy to replenish the atmospheric losses and waves propagating upward may carry these energy fluxes with them into the upper solar atmosphere. Nature of waves i.e., either longitudinal or transverse, frequency, magnetic topology, wave amplitude, propagation speed and other parameters such as plasma density, field strength, etc. may decide upon the amount of energy fluxes carried by them. 

One of the outward propagating waves is slow magnetoacoustic waves (SMAWs) which are easily detected in the imaging observations due to their compressive nature \citep[e.g.,][]{2002SoPh..209...61D}. \citet{2011SSRv..158..267B,2021SSRv..217...76B} provide comprehensive reviews on the propagation of slow magnetoacoustic waves along open structures. Slow magnetoacoustic waves are subjected to damping while propagating in the solar corona \citep[e.g.,][]{2014A&A...568A..96G,2014ApJ...789..118K}. Propagating waves develop into shocks due to amplitude steepening at heights where a sharp change in density or temperature occurs (such as chromosphere and transition region) which can be observed as saw-tooth patterns in the velocity oscillations \citep{2000SoPh..192..373B}. \citet{2000ApJ...533.1071O} found that slow magnetoacoustic waves steepen non-linearly while propagating into the corona, leading to wave-enhanced dissipation. However, slow waves shock dissipation within the sunspot umbra do not contribute to the umbral atmospheric heating \citep[e.g.,][]{1981A&A....97..310M,2011ApJ...735...65F}. Although damping of slow magnetoacoustic waves in the solar atmosphere is very well studied theoretically, calculating damping flux along the solar atmosphere observationally through wave propagation is very complex. This is due to the fact that the dynamics of the photosphere, chromosphere and corona are very different. Furthermore, in the lower atmosphere, effects due to opacity and area divergence are difficult to evaluate \citep{2016ApJ...831...24K}.

Damping of these waves are studied in various atmospheric structures including magnetic pores \citep[e.g.,][]{2021RSPTA.37900172G}, umbra \citep[e.g.,][]{2016ApJ...831...24K,2017ApJ...847....5K}, coronal holes \citep[e.g.,][]{2014A&A...568A..96G}, coronal loops \citep[e.g.,][]{2003A&A...408..755D} etc. These waves are also important for seismological studies and provide estimates on important plasma parameters that are otherwise difficult to calculate in the solar atmosphere such as adiabatic index, compressive viscosity, thermal conductivity, etc. \citep[e.g.,][]{2011ApJ...727L..32V,2015ApJ...811L..13W,2018ApJ...868..149K}. 

The power spectra of intensity and velocity oscillations in the umbral atmosphere reveal the presence of enhanced powers in various period bands \citep[e.g.,][]{2017ApJ...847....5K}. In sunspots at the photosphere, the power of 5-min oscillations dominates over 3-min oscillations \citep{2000SoPh..192..373B}. At heights above the photosphere, 3-min oscillations dominate due to the acoustic cut-off of 5-min oscillations \citep[e.g.,][]{1991A&A...250..235F,2016PhDT........15L}. The presence of 3-min oscillations at specific locations in the umbral photosphere was reported by \citet{2012ApJ...757..160J}, and their propagation from photosphere to corona along fan loops was studied in detail by \citet{2023MNRAS.525.4815R,2024BSRSL..93..948R}. Moreover, low period oscillations ($\approx$ 1-min) at some atmospheric heights in the umbra are also reported \citep[e.g.,][]{2017ApJ...847....5K,2020A&A...638A...6S}. \citet{2018ApJ...856L..16W} reported low period oscillations ($\approx$ 1.5 min) at all the heights above the umbra from imaging observations. \citet{2016A&A...594A.101Y} reported low-period oscillations in the corona at the location of light-bridge in sunspot. While propagating, these waves show evidence of damping. Damping of slow magnetoacoustic waves of various periods along the integrated umbral atmosphere from photosphere to transition region was studied by \citet{2017ApJ...847....5K} and along open structures in corona were studied by \citet{2014A&A...568A..96G,2014ApJ...789..118K} etc. All these results indicate some frequency-dependent damping of slow magnetoacoustic waves. Moreover, recently \citet{2024MNRAS.527.5302M} reported different damping lengths for 3-min slow waves propagating along the same fan loop observed from different instruments with non-parallel lines of sight.

\begin{figure*}
    \centering
    \includegraphics[width=0.55\textwidth]{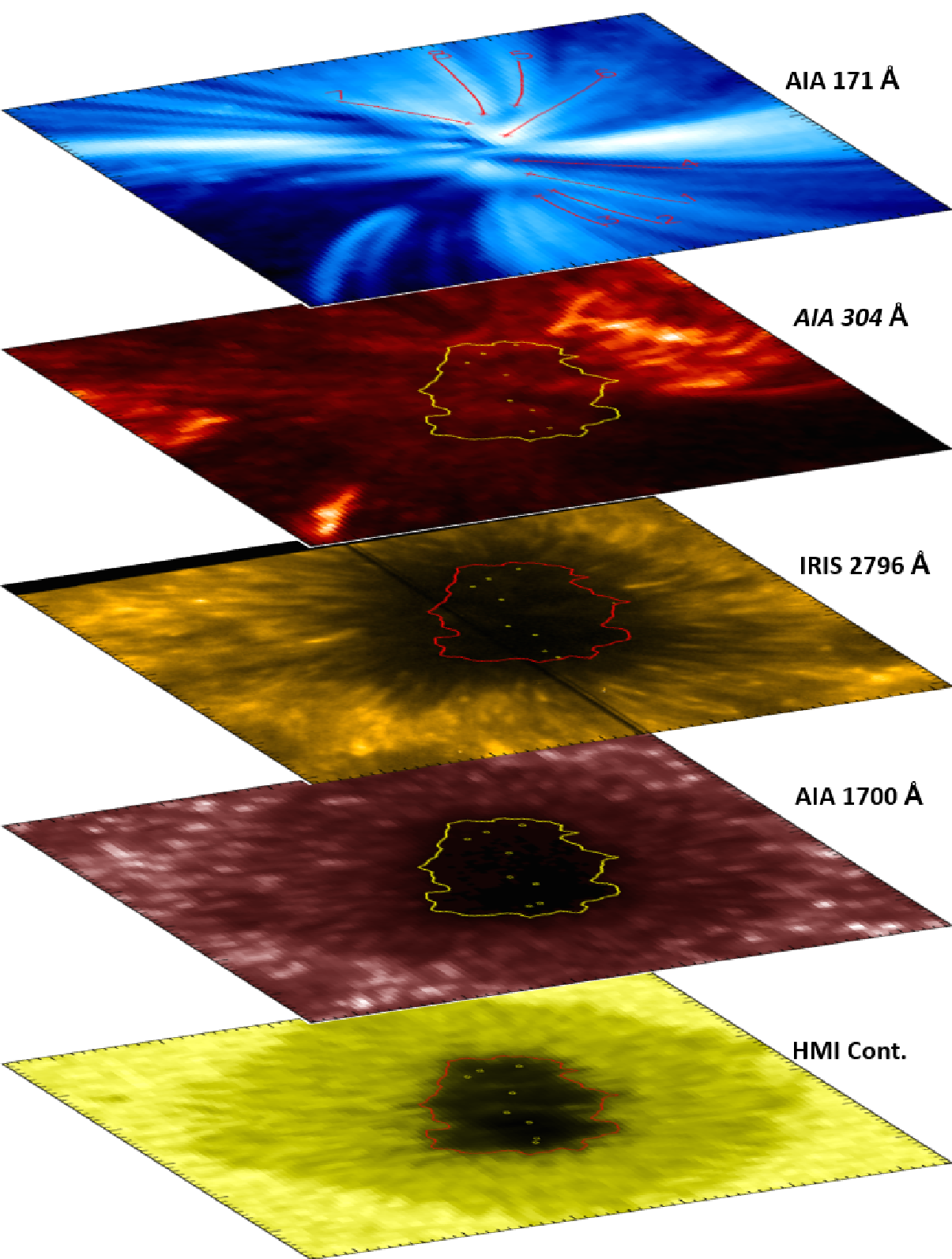}\includegraphics[width=0.4\textwidth]{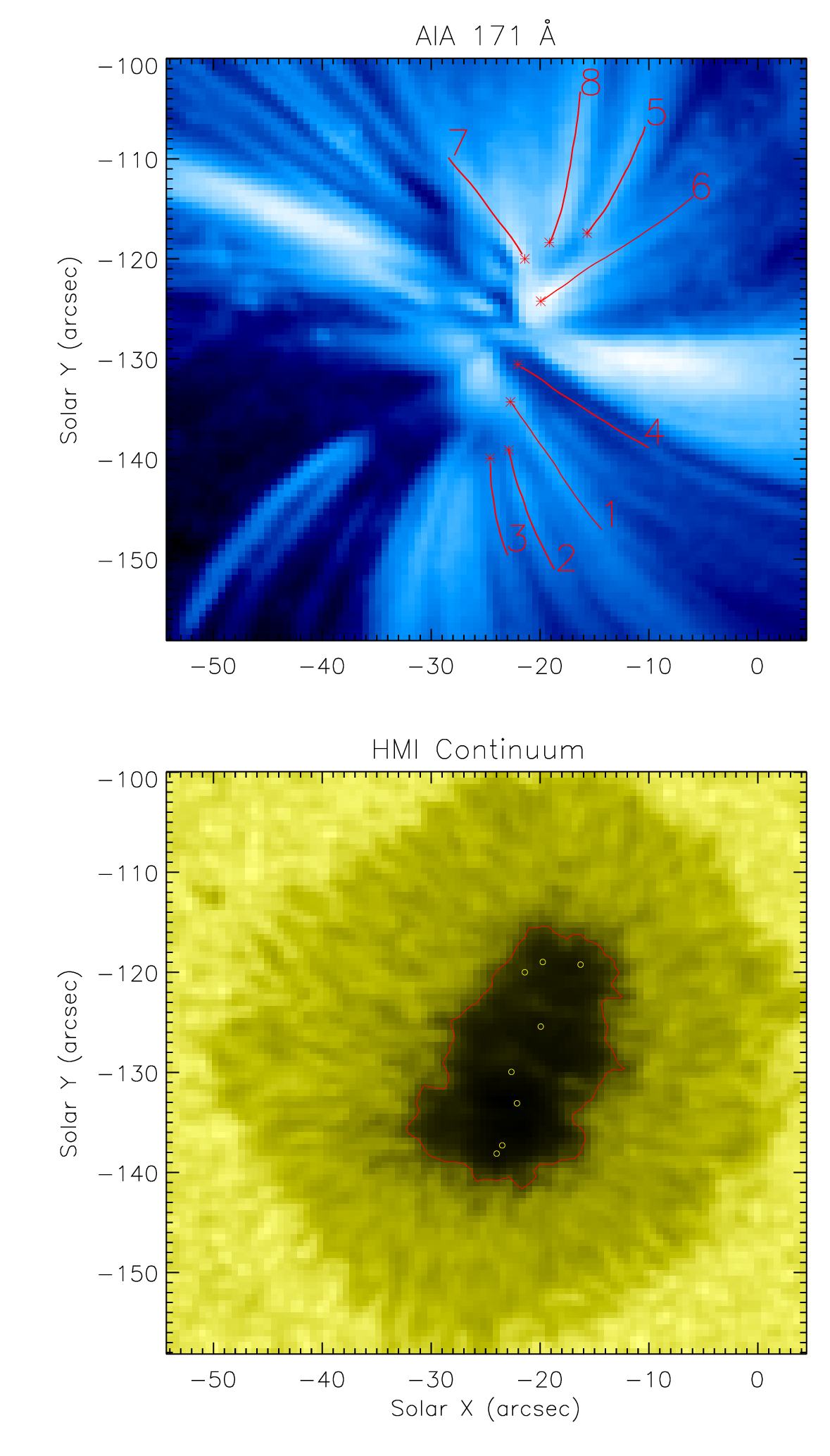}
    \caption{Images of sunspot and fan loops obtained at 7:19 UT on June 16, 2016 by various passbands as labeled. Asterisk (*) represents coronal footpoints of all the loops as identified from AIA 171 \AA\ image. Small circles (o) represent the location of all the identified loops at their respective atmospheric heights \citep[see details in][]{2023MNRAS.525.4815R}. Contours indicate the umbra-penumbra boundary as obtained from the HMI continuum image.}
    \label{fig:maps}
\end{figure*}

To balance the total horizontal pressure, the cross-sectional area of the magnetic flux tube increases with height in the solar atmosphere. This leads to the damping of slow magnetoacoustic waves due to area divergence and is related to the conservation of wave energy within the flux tube. Area divergence causes the wave amplitude to decay as $1/\sqrt{(A(s))}$, where A(s) is the cross-sectional area of coronal loop with height \citep{2004A&A...415..705D}. Area divergence does not contribute to heating as the wave energy redistributes itself to conserve wave energy within the flux tube which is purely a geometric effect. These expanding fields also enhance the transmission of waves to the corona by decreasing the reflection from chromosphere and transition region \citep{2017ApJ...840...20S}. Therefore, it is important to take into account the effect of area divergence for an accurate estimate of wave damping. 
 
Till now most of the work on slow magnetoacoustic wave energy propagation in the umbral atmosphere is carried out either over the whole integrated umbra or at some unidentified location within the umbra \citep[e.g.,][]{2016ApJ...831...24K,2017ApJ...836...18C,2017ApJ...847....5K}. In this work, our focus is to study the damping of slow magnetoacoustic waves propagating along the already identified fan loops rooted in the sunspot umbra \citep{2023MNRAS.525.4815R}. Here, we are exploring the effect of area divergence and frequency on the damping of these waves while propagating from photosphere to corona along fan loops. We describe observations in Section~\ref{sec:obs}, data analysis and results in Section~\ref{sec:analysis}, and finally summarise and discuss our results in Section~\ref{sec:discussion}.

\section{Observations}
\label{sec:obs}

To investigate the damping of slow magnetoacoustic waves in the solar atmosphere, we utilize multi-wavelength observations of fan loops within the active region NOAA AR 12553 observed on June 16, 2016 by Atmospheric Imaging Assembly \citep[AIA;][]{2012SoPh..275...17L}, Helioseismic and Magnetic Imager \citep[HMI;][]{2012SoPh..275..207S} both onboard Solar Dynamics Observatory \citep[SDO;][]{2012SoPh..275....3P}, and Interface Region Imaging Spectrograph \citep[IRIS;][] {2014SoPh..289.2733D}. We obtained 4 hours of simultaneous data starting from 07:19:13 UT. The same dataset was previously utilized by \citet{2023MNRAS.525.4815R,2024BSRSL..93..948R} to trace the source region of 3-min slow magnetoacoustic waves propagating along the fan loops rooted in the umbral region.

AIA/SDO provides full-disc images of the Sun in seven EUV and three UV-visible passbands covering the upper and lower solar atmospheres, respectively. HMI/SDO provides full-disc images of the photospheric Sun in the intensity continuum, Dopplergram and magnetogram derived from Fe I 6173 \AA\ spectral line. All the SDO images were calibrated, co-aligned, and rescaled to a common 0.6$\arcsec$ pixel$^{-1}$ resolution and 12-s temporal resolution using the robust SDO library of Rob Rutten\footnote{https://robrutten.nl/rridl/00-README/sdo-manual.html} which incorporates spline interpolation to rescale the time series. Images obtained from IRIS-SJI 2796 \AA\ passband have an exposure time of 2 s with an effective cadence of 6.88 s and 0.166$\arcsec$ pixel$^{-1}$ resolution. IRIS has a field of view (FOV) of $60\arcsec \times65\arcsec$ which is centered around the sunspot umbra. Details on data preparation are provided in \citet{2023MNRAS.525.4815R}. Fig.~\ref{fig:maps} shows the images of the sunspot and fan loops as obtained from AIA, IRIS and HMI passbands as labeled. 

The selected sunspot is slightly off the disc center (heliocentric coordinates X $\approx -25\arcsec$, Y $\approx -125\arcsec$), the angle between local vertical and line-of-sight (LOS) is $\approx 7.7 ^o$, which leads to $\mu$ = cos $\theta \approx$ 0.99. Therefore, any projection effect with respect to the disc center will be negligible. 

\subsection{IRIS spectroscopic observations}
\label{iris}

\begin{figure*}
    \centering
    \includegraphics[width=0.35\textwidth]{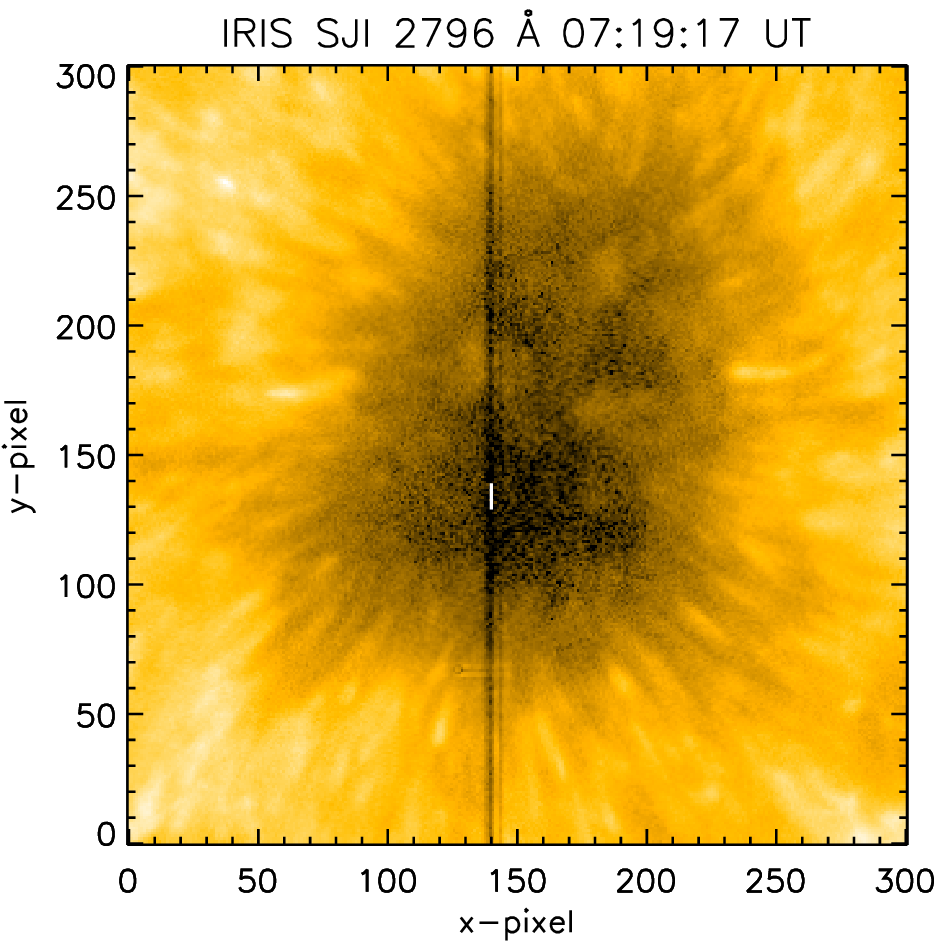}\includegraphics[width=0.64\textwidth]{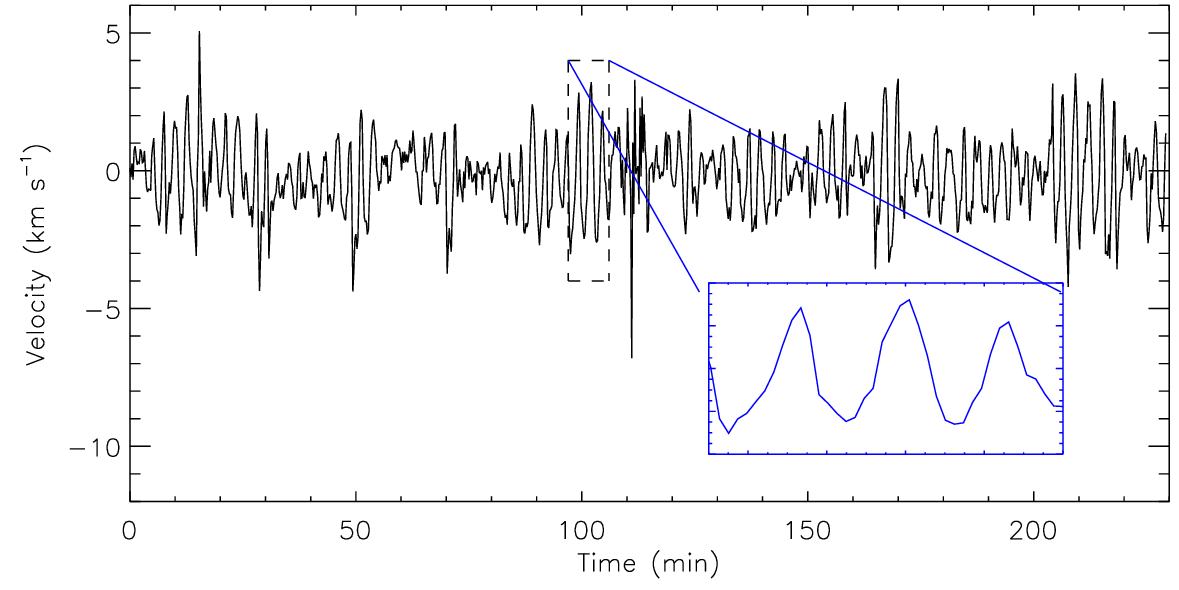}
    \caption{Left: IRIS slit-jaw image in Mg II 2796 \AA\ passband. The black line is the position of the IRIS slit. Pixels along a small white line on the slit are averaged over to determine the amplitude of velocity oscillations in the chromosphere above the sunspot umbra. Right: Temporal evolution of Doppler velocity derived from Mg II 2796 \AA\ line formed at the chromosphere above sunspot umbra. Redshifts and blueshifts of chromospheric oscillations are visible and shown by positive and negative values of velocities, respectively. The blue box shows the enlarged view of the dashed box.}
    \label{fig:iris}
\end{figure*}

IRIS also observed in spectroscopic mode and recorded the time series in 2-step raster mode. Both slit width and pixel size along the slit direction are 0.166$\arcsec$/pixel. Each raster step has an exposure time of 2 s and slit length of 399 pixels. We have extracted the near-ultraviolet spectra of Mg II at 2796 \AA\ (formation temperature of 10,000 K) from one slit location of the raster to generate time series. We averaged over the spectra for a region marked with a white line (11-pixels) in Fig.~\ref{fig:iris} and further binned over two time-frames to improve upon the signal. The extracted region is representative of the quiet umbral atmosphere. After all binning, the effective cadence becomes 13.76 s. We then fitted average spectral profiles with a single Gaussian function and constant background and further extracted the Doppler shifts and peak intensity of the profiles with time (for more details see Appendix~\ref{appendix:spectra}). The final Doppler velocity oscillations are plotted in the right panel of Fig.~\ref{fig:iris} with the blue box showing an enlarged view of the oscillatory saw-tooth pattern with time.

\section{Data Analysis and Results}
\label{sec:analysis}

Fig.~\ref{fig:maps} shows the analyzed fan loop structures in AIA 171 \AA\ passband. Overplotted contour represents the umbral boundary of a sunspot at 9000 DN, as obtained from the HMI continuum image. We have identified eight fan loops emanating from the sunspot umbra for our study purpose which are labeled accordingly. In AIA 171 \AA\ image asterisks (*) represent the coronal foot-points of fan loops from where the loops are visible in corona. Locations of all the loops at different atmospheric heights from the transition region to the photosphere are already identified \citep[see details in][]{2023MNRAS.525.4815R} and marked with small circles (o) in Fig.~\ref{fig:maps}. We estimate the inclination angle of the traced loops from the photosphere to corona to be $\approx 10^o-30^o$  as calculated from the spatial offsets between the loop locations identified at the photosphere (HMI continuum) and at the corona (AIA 171 \AA ). Here, we are presenting results from loop 6 as a representative example and present results from all the loops in Section~\ref{sec:damping}.

We carried out analysis for three different cases. Firstly, analysis is carried out for the loop locations identified at different atmospheric heights denoted by small circles (o) and asterisks (*) in Fig.~\ref{fig:maps}. These are the locations where the correlation value obtained between light curves from two atmospheric heights is maximum and represents the center of the loop. Secondly, we also carried out analysis over all the pixels that fall within the contour area where the correlation value between two atmospheric heights becomes $\approx$95\% of the maximum correlation value. This area can be considered as the upper limit on the size of the loop i.e., loop cross-section. The area within this correlation threshold decreases as we move into the lower atmosphere (see Appendix~\ref{appendix:area} for details). This is as per the expectations from the theory of flux tube expansion with height. Details on technique and methodology are described in \citet{2023MNRAS.525.4815R}. Thirdly, analysis is also carried out over the whole integrated umbra as marked by the overplotted contours in Fig.~\ref{fig:maps} for comparison with previous results \citep[e.g.,][]{2017ApJ...847....5K,2016ApJ...831...24K}.

\subsection{Fourier powers and oscillation amplitudes}

\begin{figure*}
   \centering
   \includegraphics[width=0.75\textwidth]{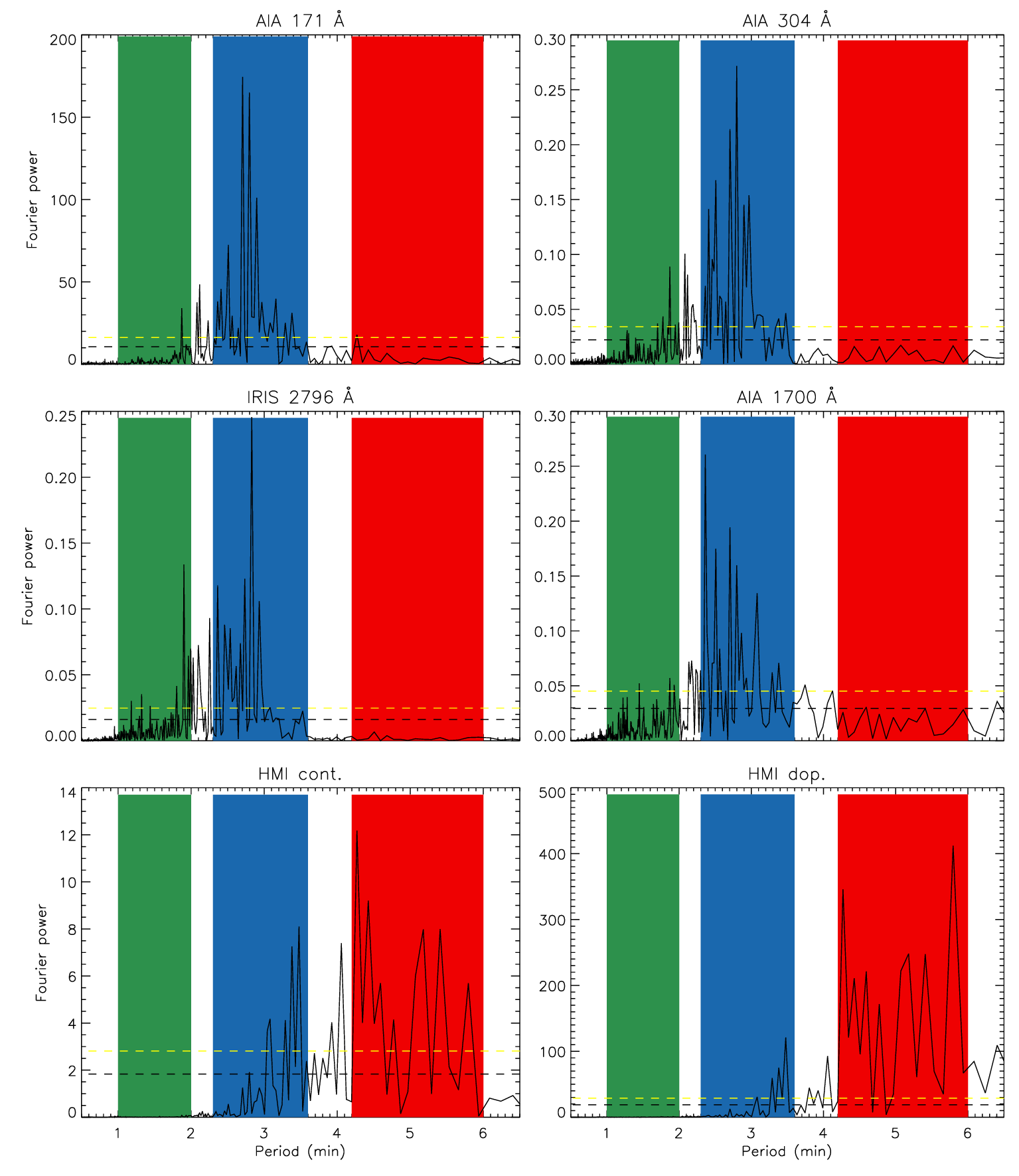}
   \caption{Fourier power spectra obtained from background subtracted light curves of loop 6 locations identified in Fig.~\ref{fig:maps} from different passbands as labeled. Shaded regions in green, blue and red colors denote 1.5-min, 3-min and 5-min period bands identified for our analysis. Horizontal dashed lines in black and yellow colors represent 95\% and 99\% confidence levels.}
    \label{fig:fourier}
\end{figure*}

\begin{figure*}
    \centering
    \includegraphics[width=0.80\textwidth]{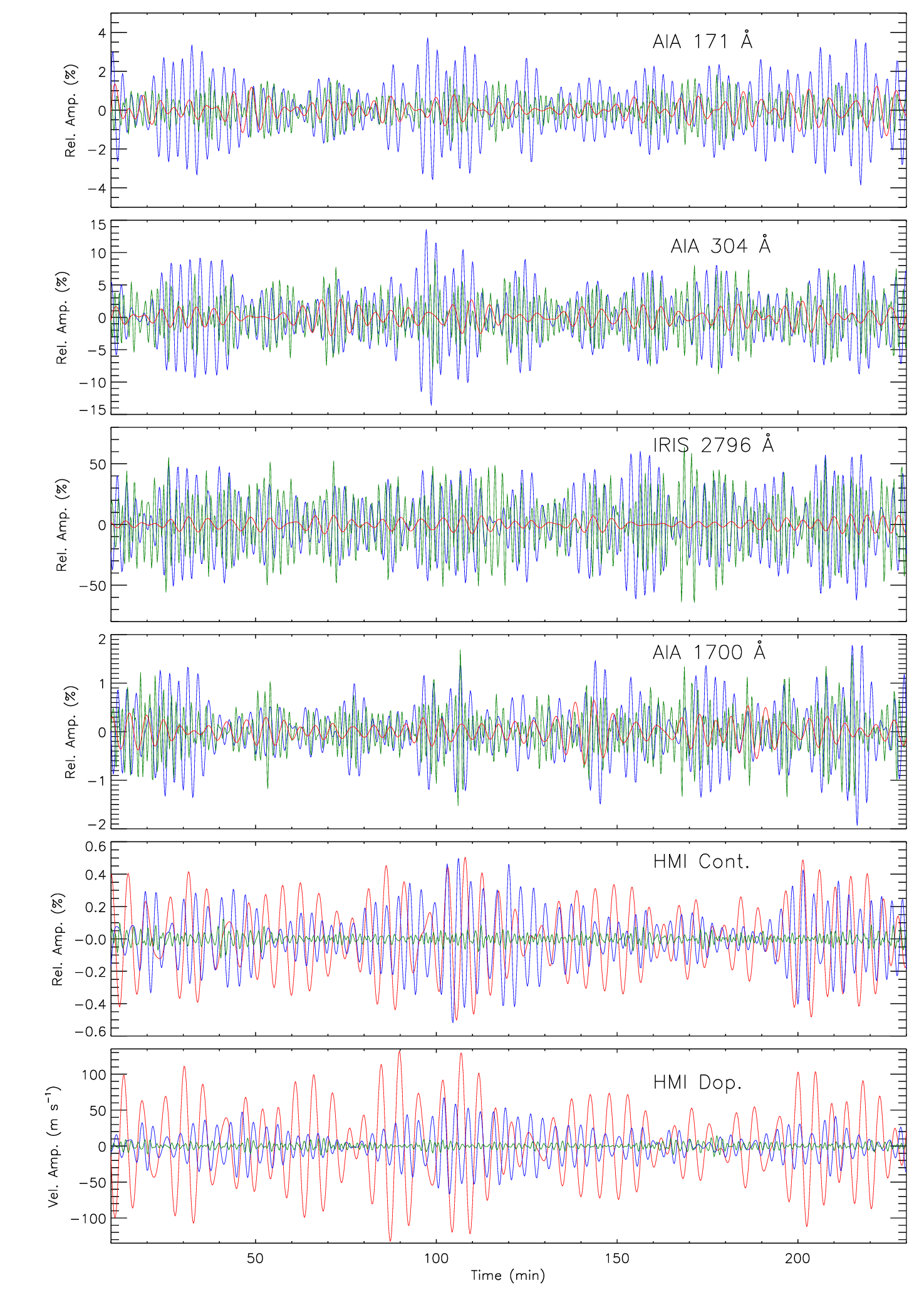}
    \caption{Relative intensity and velocity amplitudes for oscillations of 1.5-min, 3-min and 5-min periods obtained at loop locations from different passbands as labeled. Here, green, blue and red colors show 1.5-min, 3-min and 5-min period oscillations respectively. }
    \label{fig:filter}
\end{figure*}

In Fig.~\ref{fig:fourier}, we show fast Fourier transform (FFT) power spectrum obtained from the background subtracted light curves of loop location 6 from different passbands as labeled \citep[for details see][]{2023MNRAS.525.4815R}. Background light curves are obtained by taking 16-min running average of original light curves at those atmospheric heights. The FFT power spectrum and confidence levels are obtained using standard IDL routine $fft\_powerspectrum$. Plots clearly show a wide distribution of power peaks in different period ranges. 1--2 min (16.67--8.33 mHz), 2.3--3.6 min (7.25--4.63 mHz) and 4.2--6.0 min (3.97--2.78 mHz) ranges show enhanced powers at different atmospheric heights. Here, we term these three dominant period ranges as 1.5-min, 3-min and 5-min period bands respectively. It should also be noted that these period bands have different period (frequency) widths. These period bands are shaded with green, blue and red colors, respectively. We will be following the same color code to represent different period bands further in the paper. We finally sum over the FFT powers within their respective bands. Fourier power in 3-min period band is present in all the passbands which suggests the propagation of 3-min waves from the photosphere to corona along the fan loops as reported in \citet{2023MNRAS.525.4815R}. Enhanced FFT powers in the 1.5-min period band are present in AIA 1700, IRIS 2796, AIA 304 and AIA 171 \AA\ passbands. The lack of Fourier powers in this band at photospheric heights (HMI continuum and Dopplergram) could be either due to their real absence at these heights or due to poor cadence of HMI observations which is 45 s. Enhanced Fourier powers in the 5-min band are present only at the photospheric heights (HMI continuum and Dopplergram) and are absent at higher atmospheric heights which could be due to atmospheric cutoff. 

\begin{figure*}
    \centering
    \includegraphics[width=0.95\textwidth]{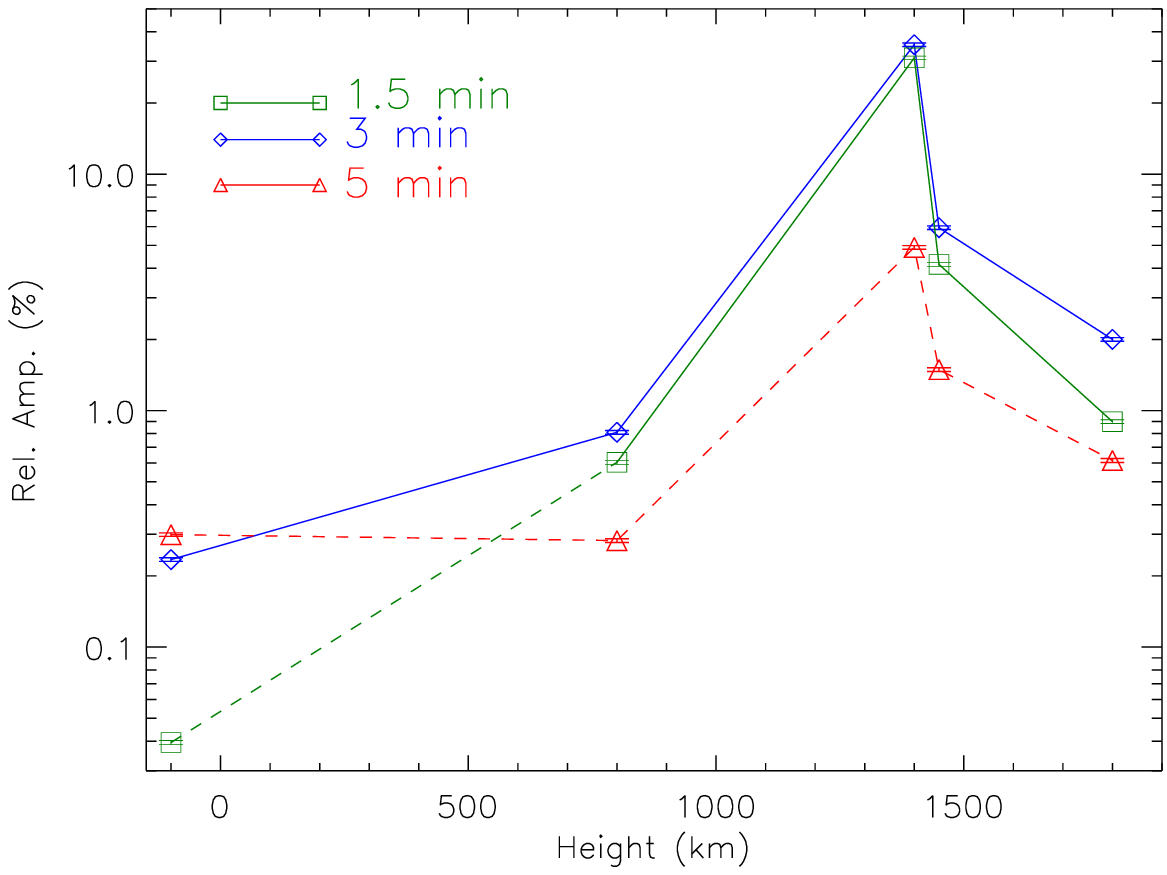}
    \caption{Variation of relative intensity amplitudes (left panel) and velocity amplitudes (right panel) of oscillations obtained at loop locations with atmospheric heights.  }
    \label{fig:maxco}
\end{figure*}

 Since we have found enhanced powers in the 1.5-min and 3-min period bands in almost all the passbands, and thus at all the atmospheric heights, we will be utilizing these period bands for further analysis of wave damping. We apply bandpass filters over these three period ranges on the original light curves (with $3\times3$ pixel binning for SDO and $5\times5$ pixel binning for IRIS). Since the absolute wave amplitudes of filtered light curves in the given period ranges can not be compared directly, these filtered light curves are normalized by background light curves present at that atmospheric heights. These normalized filtered light curves represent relative wave amplitudes with time are shown in Fig.~\ref{fig:filter} for 1.5-min, 3-min and 5-min period bands at all atmospheric heights. These curves show clean intensity oscillations in all three period bands. However, due to the presence of several nearby power peaks in the different period bands, we notice oscillations in the form of unclean wave packets. The bottom panel of Fig.~\ref{fig:filter} shows the filtered velocity oscillations obtained from HMI Dopplergram which provides the velocity amplitude variations of 1.5-min, 3-min and 5-min period bands with time. From the plots in Figs.~\ref{fig:fourier} and \ref{fig:filter}, we notice that 5-min oscillation dominates at the photosphere and 3-min oscillation dominates over 1.5-min oscillation at the higher heights where 5-min oscillation is absent. The root mean square (RMS) relative amplitudes are obtained using the formula $\sqrt{\frac{1}{n}\sum{(\frac{I'}{I_o})^2}}$ for each period band, where $I_o$ is background light curves and I' is filtered light curves obtained in the ranges shown by shaded region in Fig.~\ref{fig:fourier}. The standard errors in the RMS values are obtained using the formula $\frac{\sigma}{\sqrt{N}}$, where $\sigma$ is the standard deviation and N is the number of data points \citep{1969drea.book.....B}. The obtained RMS amplitudes and standard errors at each height are multiplied with $\sqrt{2}$ to determine the relative intensity amplitudes and associated error-bars. In the left panel of Fig.~\ref{fig:maxco}, we plot relative amplitude of intensity oscillations for different period bands present at various atmospheric heights.  
 Furthermore, we determine the relative intensity amplitude at all the pixels within the loop contours and then average them to determine the relative amplitudes of these waves within the loop cross-sections along the loop. 
 
\begin{figure*}
    \centering
    \includegraphics[width=0.95\textwidth]{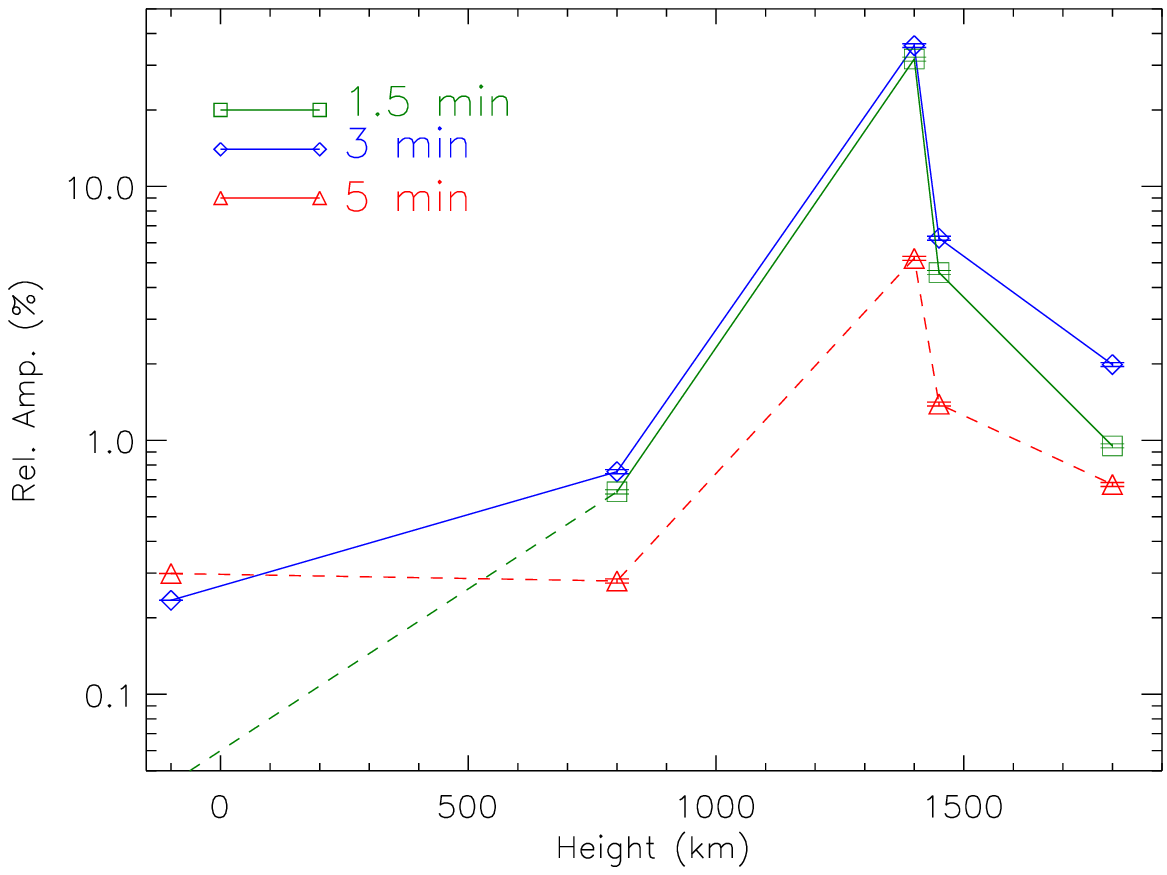} 
    \caption{ Variation of average relative intensity amplitudes (left panel) and velocity amplitudes (right panel) of oscillations obtained within the loop cross-sectional areas with atmospheric heights.} 
    \label{fig:95}
\end{figure*}

 Relative intensity amplitude variations with height are determined using different methods as mentioned earlier and are plotted in the left panel of Figs.~\ref{fig:maxco} and~\ref{fig:95} for single maximum correlated pixel and loop cross-sectional area respectively. Variations in amplitudes with height are similar for both the methods. Relative intensity amplitude variation with height initially shows amplitude steepening due to rapid fall of density. The relative intensity amplitude is maximum at chromospheric height (IRIS 2796 \AA) which makes the waves non-linear. This can also be seen in the saw-tooth pattern of velocity oscillations obtained at chromospheric height using Mg II 2796 \AA\ line as shown in the right panel of Fig.~\ref{fig:iris}, which eventually develops into shocks \citep{2006ApJ...640.1153C}. High-period (5-min) shows atmospheric cutoff as the amplitude of these waves decreases as they move from photospheric height to chromospheric height. As seen in the FFT power spectrum, enhanced power in 1.5-min period band is absent only at photosphere whereas enhanced power in 5-min period band is present only at photosphere. Therefore, the relative intensity amplitudes of 5-min period at heights above photosphere and 1.5-min period at photosphere are unreliable. Henceforth, all these unreliable data points are connected using dashed lines in Fig.~\ref{fig:maxco} and further for representation purpose only. 
 
Velocity amplitudes of 1.5-min, 3-min and 5-min oscillations are determined assuming linear approximation in transition region (AIA 304 \AA) and corona (AIA 171 \AA) using formula
\begin{math}
    \frac{I'}{2I_o} = \frac{v'}{v_s},
\end{math}
where $v'$ is the velocity amplitude and $v_s$ is the speed of sound \citep{2009A&A...503L..25W}. We determined the amplitude of velocity oscillations at the photosphere using HMI Dopplergram data for 3-min and 5-min to be $\approx$ 33 m s$^{-1}$ and 73 m s$^{-1}$, respectively as shown in Fig.~\ref{fig:filter}. For chromospheric height, velocity amplitude of 1 km s$^{-1}$ was found for 3-min oscillations from Mg II 2796 \AA\ line (see details in Appendix~\ref{appendix:spectra}). Using the velocity amplitudes of photospheric and chromospheric oscillations, we linearly interpolated and extracted the velocity amplitude of 3-min oscillations for temperature minimum region.

These derived velocity amplitudes of 3-min oscillations will be utilized to calculate wave energy flux. Since we have relative intensity amplitude of 1.5-min, 3-min and 5-min intensity oscillations, so to calculate velocity amplitudes of 1.5-min and 5-min oscillations for temperature minimum and chromospheric heights, we scaled their relative intensity amplitudes with relative intensity and velocity amplitudes of 3-min oscillations at respective atmospheric heights. Finally, the obtained velocity amplitude variations with height are shown in the right panel of Figs.~\ref{fig:maxco} and \ref{fig:95} for maximum correlated pixel and loop cross-sectional area respectively. Moreover, inclination of loop from photosphere to corona will change the velocity perturbations by inverse factor of cos($10^o-30^o$) $\approx$ 0.98-0.87 (see Section~\ref{sec:analysis}). This will lead to changes of only about 2-15\% and are within the estimated error bars of velocity amplitude.

\subsection{Energy flux of waves} 
\label{sec:flux}

\begin{figure*}
    \centering
    \includegraphics[width=0.95\textwidth]{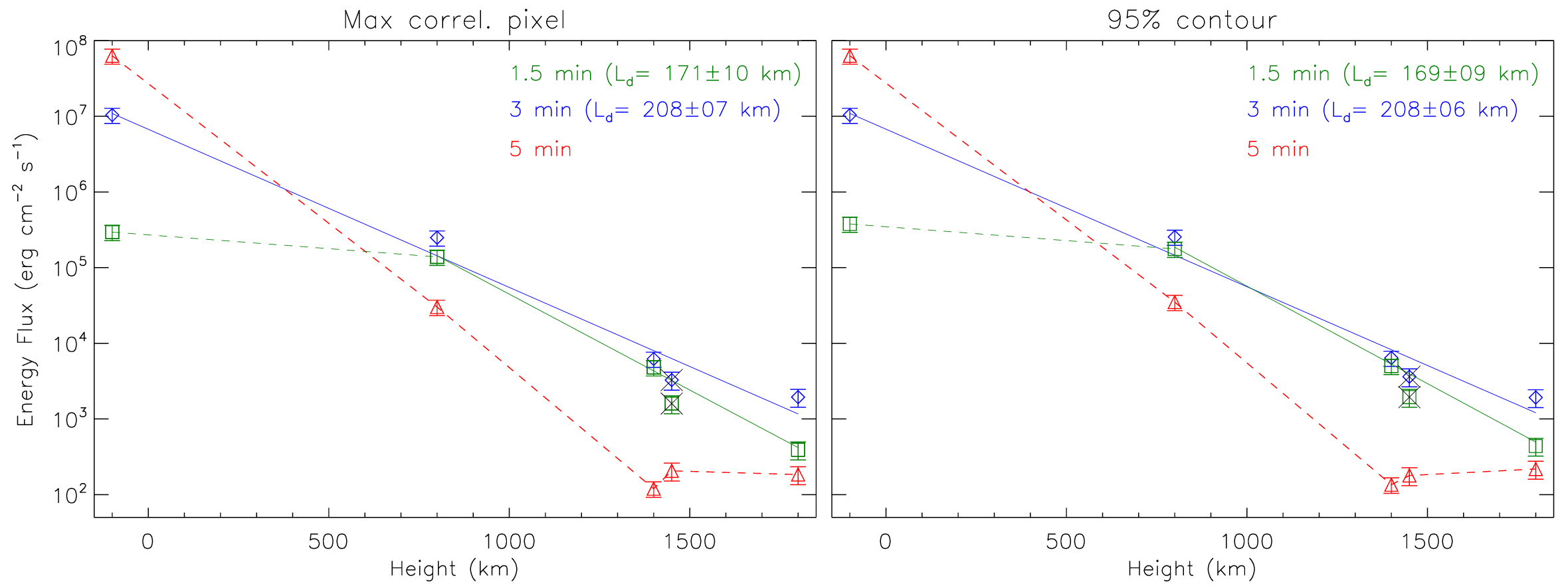}
    \caption{Variation of wave energy ﬂuxes at loop locations (left panel) and average wave energy fluxes within the loop cross-sectional areas (right panel) with atmospheric heights. }
    \label{fig:fluxcontour}
\end{figure*}

To calculate the wave energy flux (F), we are using the WKB approximation, $F= \rho \delta v^2 v_p $ where $\rho, \delta v, v_p$, are mass density, velocity amplitude, and slow wave propagation speed respectively. However, under a typical coronal condition of low-beta plasma, sound speed ($v_s$) is a good approximation to the phase speed of slow magnetoacoustic waves \citep{2009A&A...503L..25W}. Since $\mu \approx$ 0.99 for our analyzed sunspot, we can assume $v_p$ equals $v_s$ and also any projection effect on velocity perturbations will also be negligible and within the error bars. Sound speed is calculated using the formula $v_s = \sqrt{ \frac{\gamma R T} {\mu}} $, where $\gamma= \frac{5}{3}$ is the polytropic index, R =$8.314\times10^7$ erg K$^{-1}$ mol$^{-1}$ is the gas constant, $\mu$ is the mean molecular weight, and T is the temperature \citep{2004psci.book.....A}. Further, we are utilizing density and temperature from the sunspot atmospheric model of \citet{1999ApJ...518..480F} and assume 20\% errors in density and temperature, see details in Appendix~\ref{appendix:temp-dens}.  

The estimated wave energy fluxes are decreasing from $\approx 10^7$ erg cm$^{-2}$ s$^{-1}$ at the photosphere to $\approx 2\times 10^3$ erg cm$^{-2}$ s$^{-1}$ at the low corona for 3-min waves. For 1.5 min waves, energy flux is $\approx 3\times 10^5$ erg cm$^{-2}$ s$^{-1}$ at the temperature minimum which decreases to $\approx 400$ erg cm$^{-2}$ s$^{-1}$ at the low corona. Wave energy flux of 1.5-min waves at the photosphere may be unreliable due to the poor cadence of the HMI continuum. For 5-min waves, energy flux at photosphere is $\approx 6\times 10^7$ erg cm$^{-2}$ s$^{-1}$ which decays rapidly in the atmosphere. These variations in wave energy fluxes are plotted in Fig.~\ref{fig:fluxcontour} for loop center locations at different heights and also for average wave energy fluxes within the loop cross-sectional areas at 95\% contour level. Plots show consistent decrease in wave energy fluxes with height even though the oscillation amplitudes are increasing with height in the lower atmosphere due to rapid fall in density. Decrease in wave energy fluxes with height indicates the damping of slow magnetoacoustic waves across these atmospheric layers. 

For comparison purpose, we also provide amplitude of oscillations and average wave energy fluxes over the whole integrated umbra with height in Appendix~\ref{appendix:umbra}. In this case also obtained amplitudes and wave energy fluxes at different atmospheric heights are of similar order of magnitude as those obtained for previous cases.

\subsection{Dependence of wave damping on frequency and area divergence}
\label{sec:damping}

The wave energy flux at different heights in the solar atmosphere are estimated for different cases as mentioned in Section~\ref{sec:flux} and plotted in Fig.~\ref{fig:fluxcontour}. To determine the dependence of frequency and area divergence on the damping of these waves, we fitted the obtained energy fluxes using exponential decay function. In Fig.~\ref{fig:fluxcontour}, solid lines represent the best fits with the data following a function $F={F}_{0}{e}^{-\frac{h}{{L}_{d}}}$, where F is the wave energy flux with height h (see  Appendix~\ref{appendix:temp-dens}), $L_d$ is the damping length, and $F_0$ is appropriate constant. Fits are performed using MPFITFUN routine \citep{2009ASPC..411..251M}. The exponential decay function appears to be consistent with the data. To improve the chi-square of fit, points in black color cross as shown in Fig.~\ref{fig:fluxcontour} are excluded from the fit. Damping lengths ($L_d$) obtained from the maximum correlated pixel are $\approx$ 171$\pm$10 km and 208$\pm$07 km whereas those from loop cross-sectional area are 169$\pm$09 km and 208$\pm$06 km for 1.5-min and 3-min period waves respectively. Damping lengths are also printed in Fig.~\ref{fig:fluxcontour}. The damping length of 1.5-min period wave is less than 3-min period wave. This indicates some frequency-dependent damping of slow waves where high-frequency waves are damped faster than low-frequency waves. Also, damping lengths obtained from maximum correlated pixels and loop areas are similar. This may indicate that wave energy fluxes are uniformly distributed across the loop cross-sections.

\begin{figure}
    \centering
    \includegraphics[width=0.5\textwidth]{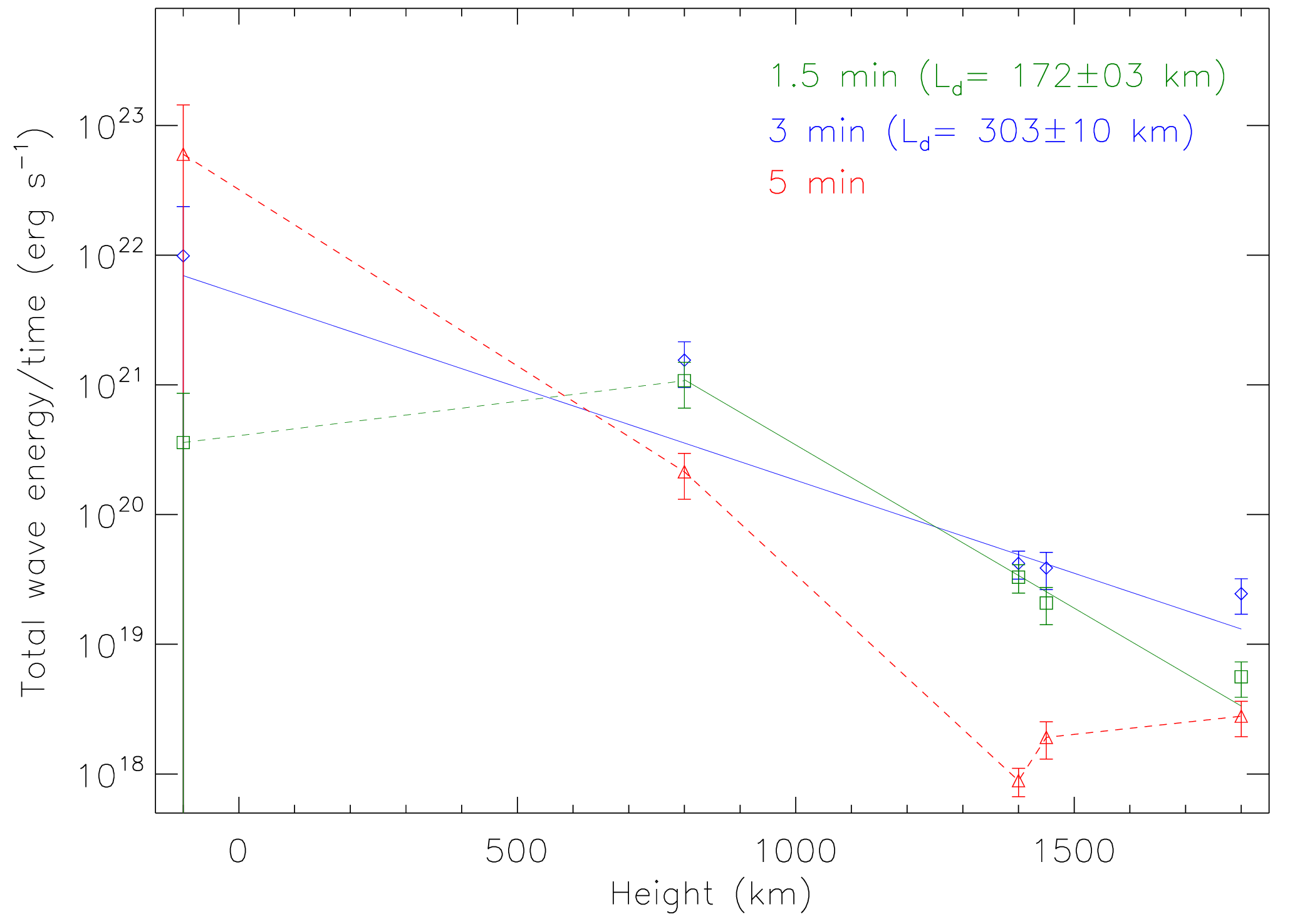}
    \caption{Variation of total wave energy content within the loop cross-sectional areas with atmospheric heights.}
    \label{fig:powercontour}
\end{figure}

Since loops are expanding with the height as reported in \citet{2023MNRAS.525.4815R}, the wave energy is redistributed over the whole loop cross-section to conserve the wave energy within the flux tube. This effect will lead to the decay of wave energy flux with height which is purely a geometric effect without any actual physical damping of waves. To take this geometric effect into account so as to estimate the actual damping of slow magnetoacoustic waves, we multiplied the wave energy flux with the cross-sectional area of the loop at that respective height. The cross-sectional area of the loop at $\approx$ 95 \%  contour level is provided in Appendix~\ref{appendix:area}. The obtained variation of total wave energy content at different atmospheric heights for different periods are shown in Fig.~\ref{fig:powercontour}. The plot clearly demonstrates the actual decay of total wave energy content carried by slow magnetoacoustic waves propagating from photosphere to corona along the fan loop. We again fitted the total wave energy curves with exponential decay functions to obtain the actual damping lengths. Obtained actual damping lengths ($L_d$) in this case are $\approx$ 172$\pm$03 km and 303$\pm$10 km for 1.5-min and 3-min period waves respectively. These damping lengths are for actual damping of total wave energy content with height which are larger than the damping lengths obtained from the wave energy fluxes with height. Findings indicate that area divergence does play a role in the damping of slow magnetoacoustic waves. 

We performed the same analysis on the other seven loops identified in Fig.~\ref{fig:maps}. Here also all the loops show the presence of 1.5-min, 3-min and 5-min period waves. Similar variations of relative intensity and velocity amplitudes along the solar atmosphere are noted. Damping lengths of wave energy fluxes for 1.5-min period are less than the damping lengths of 3-min period for all the loops. Damping lengths obtained from identified loop locations and loop cross-sectional areas are listed in Table~\ref{tab:1}. Findings confirm almost uniform distribution of wave energy fluxes across the loop cross-sections. However, the longer damping length of the total wave energy content propagating along the different loops is due to area expansion of the loops, and thus indicates area-dependent damping of waves. Table~\ref{tab:1} also shows different damping lengths for 1.5-min and 3-min period bands as found above. Results provide a clear dependence of frequency and loop area divergence on the damping of slow waves.

\begin{table*}
    \centering
        \caption{Damping lengths (km) for slow magnetoacoustic wave energy flux (erg cm$^{-2}$ s$^{-1}$) and total wave energy content (erg s$^{-1}$) propagating along different fan loops identified in Fig.~\ref{fig:maps}. Damping lengths (km) are presented for loop locations and within loop cross-sectional areas obtained at 95\% contour level.}
   \label{tab:1}
   { \begin{tabular}{|c|c|c|c|c|c|c|} 
    \hline
    
\multicolumn{1}{|c|}{}   &  
\multicolumn{2}{|c|}{Damping length (km) for }   &

\multicolumn{2}{|c|}{Damping length (km) for average } &
\multicolumn{2}{|c|}{Damping length (km) for total wave  }  \\

\multicolumn{1}{|c|}{Loop no.}   &  
\multicolumn{2}{|c|}{ flux at max correl. pixel }   &

\multicolumn{2}{|c|}{flux for pixels within 95\% contour} &
\multicolumn{2}{|c|}{energy content within 95\% contour}  \\ 
\hline
  & for 1.5-min & for 3-min & for 1.5-min & for 3-min & for 1.5-min & for 3-min \\  \hline
1 &131$\pm$07 &190$\pm$05&139$\pm$06&188$\pm$05&168$\pm$04&231$\pm$06     \\ \hline
2 &161$\pm$14 &307$\pm$18&165$\pm$14&259$\pm$13&160$\pm$04&250$\pm$11    \\ \hline
3 &169$\pm$15 &243$\pm$11&160$\pm$13&241$\pm$11&129$\pm$03&225$\pm$10   \\ \hline
4 &170$\pm$10 &220$\pm$08&175$\pm$10&226$\pm$06&186$\pm$04&300$\pm$04    \\ \hline
5 &231$\pm$24 &220$\pm$13&222$\pm$21&216$\pm$09&250$\pm$04&341$\pm$11   \\ \hline
6 &171$\pm$10 &208$\pm$07&169$\pm$09&208$\pm$06&172$\pm$03&303$\pm$10   \\ \hline
7 &155$\pm$11 &220$\pm$08&166$\pm$12&221$\pm$08&167$\pm$04&264$\pm$09   \\ \hline
8 &199$\pm$18 &249$\pm$12&195$\pm$16&246$\pm$11&203$\pm$04&306$\pm$10   \\ \hline

\end{tabular}}
 \end{table*}

\section{Discussion and Summary}
\label{sec:discussion}

In this work, we utilized FFT power spectra of fan loops rooted in sunspot umbra and found waves with different periods. We noticed enhanced powers in three period bands such as 1–-2 min (16.67--8.33 mHz), 2.3–-3.6 min (7.25--4.63 mHz) and 4.2--6.0 min (3.97--2.78 mHz) at different atmospheric heights. 1--2 min period band is present above the photosphere, 2.3--3.6 min period band is present at all the atmospheric heights whereas 4.2--6 min period band is present only at the photosphere. The 1.5-min period band is also reported by \citet{2018ApJ...856L..16W} at different atmospheric heights within the umbra. The absence of 1.5-min period at the photosphere could be either due to poor cadence of HMI data or due to their real absence in our observations. However, it should be noted that \citet{2017ApJ...847....5K} also did not find any enhancement in 1.5-min period band at the photospheric height despite having better cadence data.  \citet{2016A&A...594A.101Y} detected short period ($\approx$1-min) oscillations at light-bridge locations within sunspot umbra. Additionally, we could not find any significant power enhancement in the 3-min period band in the power spectrum of the integrated umbra at the photosphere, though significant enhancements were observed at heights above the photosphere (see Appendix~\ref{appendix:umbra}). This is similar to the previous reports \citep[e.g.,][]{2017ApJ...847....5K}. This could be due to the fact that 3-min enhancements are found only at the specific locations in the umbral photosphere potentially at the sites of umbral dots \citep{2012ApJ...757..160J} and also at the foot-points of fan loops \citep{2023MNRAS.525.4815R}. 

We estimated wave energy fluxes for 3-min and 5-min period bands at the photosphere which are $\approx 10^7$ and $\approx 6\times 10^7$ erg cm$^{-2}$ s$^{-1}$ respectively whereas that for 1.5-min period at temperature minimum region is $\approx 5\times 10^5$ erg cm$^{-2}$ s$^{-1}$ which are similar to the previous reports. \citet{2016ApJ...831...24K} estimated the slow magnetoacoustic waves energy ﬂuxes for 6-10 mHz (1.7-2.7 min) band at the photosphere as $2\times10^7$ erg cm$^{-2}$ s$^{-1}$ and \citet{2017ApJ...847....5K} reported energy ﬂuxes to be $\approx 10^7$ erg cm$^{-2}$ s$^{-1}$ for 3-min period waves. For 5-min period band, \citet{2021RSPTA.37900172G} reported slow wave energy flux to be $\approx 3 \times 10^7$ erg cm$^{-2}$ s$^{-1}$ at an atmospheric height of 100 km inside the pore. As waves propagate upward these energy fluxes decay with height and provide evidence of wave damping as demonstrated in Fig.~\ref{fig:fluxcontour}. The sharp drop in energy fluxes from photosphere to chromosphere can be due to leaky wave characteristic \citep{1986SoPh..103..277C}, strong shocks dissipation and radiative cooling \citep{2011ApJ...735...65F}. These waves propagating along loops can also get damped due to various non-ideal MHD effects depending on the different physical conditions of the loop \citep{2021SSRv..217...34W}. There are several damping mechanisms proposed for slow magnetoacoustic waves such as compressive viscosity \citep[e.g,][]{2000ApJ...533.1071O}, thermal conduction \citep[e.g,][]{2003A&A...408..755D}, area divergence \citep[e.g,][]{2004A&A...415..705D}, mode coupling \citep[e.g,][]{2004A&A...415..705D}, shocks \citep[e.g,][]{2008ApJ...685.1286V} etc. 

In this work, we obtained the damping length of $\approx$208 km for 3-min slow wave energy flux propagating along fan loops from photosphere to corona and $\approx 170$ km for 1.5-min period waves propagating from temperature minimum region to corona. \citet{2017ApJ...847....5K} reported damping lengths of $\approx 149$ km and 119 km from decay of relative intensity amplitude of slow wave propagation in integrated umbra from chromosphere to transition region for periods $\approx 2.5$-min and 1.3-min respectively. Both the results indicate some frequency-dependent damping of slow magnetoacoustic waves in the lower atmosphere. Frequency-dependent damping lengths are also reported in the corona along open structures \citep[e.g.,][]{2014A&A...568A..96G,2014ApJ...789..118K}. We also noted that wave energy flux at the center of the loop and average wave energy flux obtained within the loop cross-sections (at $\approx$ 95\% contour level) are of similar order (see Fig.~\ref{fig:fluxcontour}). Fluxes are similar because velocity amplitudes at all the pixels within the loop cross-sections are almost similar.
This may indicate that wave energy fluxes are uniformly distributed across the fan loop cross-sections and also a similar damping mechanism is operating throughout the umbra. Therefore, damping of slow magnetoacoustic waves propagating in different regions of the solar atmosphere shows some kind of frequency-dependent damping even though different damping mechanisms might be operating at different heights.  

One of the dominant damping mechanisms of slow magnetoacoustic waves is area divergence which is very well studied in the solar corona \citep{2004A&A...415..705D}. However, its influence in the lower atmosphere is still unexplored. In this work, we utilized the cross-sectional area of the fan loops obtained at different heights in the lower atmosphere \citep{2023MNRAS.525.4815R}. We explored the effect of area expansion on wave damping, by multiplying the loop cross-sectional area with the energy flux of slow waves at that atmospheric height. The obtained total slow wave energy propagating along fan loop in unit time at different heights is plotted in Fig.~\ref{fig:powercontour} and represents actual damping of slow waves. 
Here damping lengths are $172\pm03$ km and $303\pm10$ km for 1.5-min and 3-min periods respectively. After incorporating the effect of area divergence, the actual damping lengths obtained are larger than those found from the damping of wave energy fluxes. This indicates that the waves are losing energy faster (i.e., damping length is short) when the area divergence effects are not considered. This is due to the fact that wave energy is redistributed across the loop cross-sections due to area divergence. This clearly explains the importance of area divergence in the damping of waves. 

Moreover, it should be noted that calculating the actual damping of wave energy flux along the solar atmosphere is a very complex problem as the dynamics of photosphere, chromosphere and corona are very different. Furthermore, the umbral photosphere itself is a complex region due to the presence of plasma-$\beta= 1$ layer which affects the propagation of wave modes \citep[e.g.,][]{2015ApJ...807...20P}. Additionally while calculating the energy fluxes, we are assuming waves to be linear. Calculated energy fluxes for 3-min waves at photosphere, chromosphere and corona are $\approx 10^7$, $\approx 10^4$ and $\approx 10^3$ erg cm$^{-2}$ s$^{-1}$. However, how much of this wave energy flux is damped, how much is getting reflected, and how much is propagating upwards still needs detailed investigation \citep[e.g.,][]{2017ApJ...840...20S}. Detailed MHD simulations with non-local thermodynamic equilibrium modeling will be required to understand the wave dynamical processes at various atmospheric heights \citep[e.g.,][]{2023A&A...670A.133F}.

In summary, we studied the damping of slow magnetoacoustic waves propagating along fan loops in the umbral atmosphere and showed oscillations in the period bands 1.5-min, 3-min and 5-min. We estimated slow magnetoacoustic wave energy fluxes propagating along the fan loops at different heights and provided the evidence of wave damping with damping lengths of $\approx 170$ km and $\approx 208$ km for 1.5-min and 3-min periods respectively for loop 6. We showed the decay of total wave energy content within the loop cross-section with height and provided evidence of actual damping of slow magnetoacoustic waves from the photosphere to corona with damping lengths of  $\approx 172$ km and $\approx 303$ km for 1.5-min and 3-min periods respectively for the same loop. Results showed some frequency-dependent damping of slow magnetoacoustic wave energy flux with height where high-frequency waves are damped faster than low-frequency waves. We have also demonstrated the importance of the role of area divergence in the damping of slow magnetoacoustic waves. 
 
\section*{Acknowledgments}
We thank the referee for helpful comments and suggestions that improved the quality of presentation. The research work at the Physical Research Laboratory (PRL) is funded by the Department of Space, Government of India. AR thanks PRL for her PhD research fellowship. AIA and HMI data are courtesy of SDO (NASA). IRIS is a NASA small explorer mission developed and operated by LMSAL with mission operations executed at NASA Ames Research Center and major contributions to downlink communications funded by the Norwegian Space Center (NSC, Norway) through an ESA PRODEX contract.

\section*{Data Availability}

The observational data utilized in this study from AIA and HMI on-board SDO are available at   \url{http://jsoc.stanford.edu/ajax/lookdata.html} and data from IRIS mission are available at \url{https://www.lmsal.com/hek/hcr?cmd=view-recent-events\&instrument=iris}.





\bibliographystyle{mnras}
\bibliography{bibli.bib} 



\appendix
\section{Appendix}

\subsection{Area of loop cross-sections}
\label{appendix:area}

\begin{figure}
    \centering
    \includegraphics[width=0.5\textwidth]{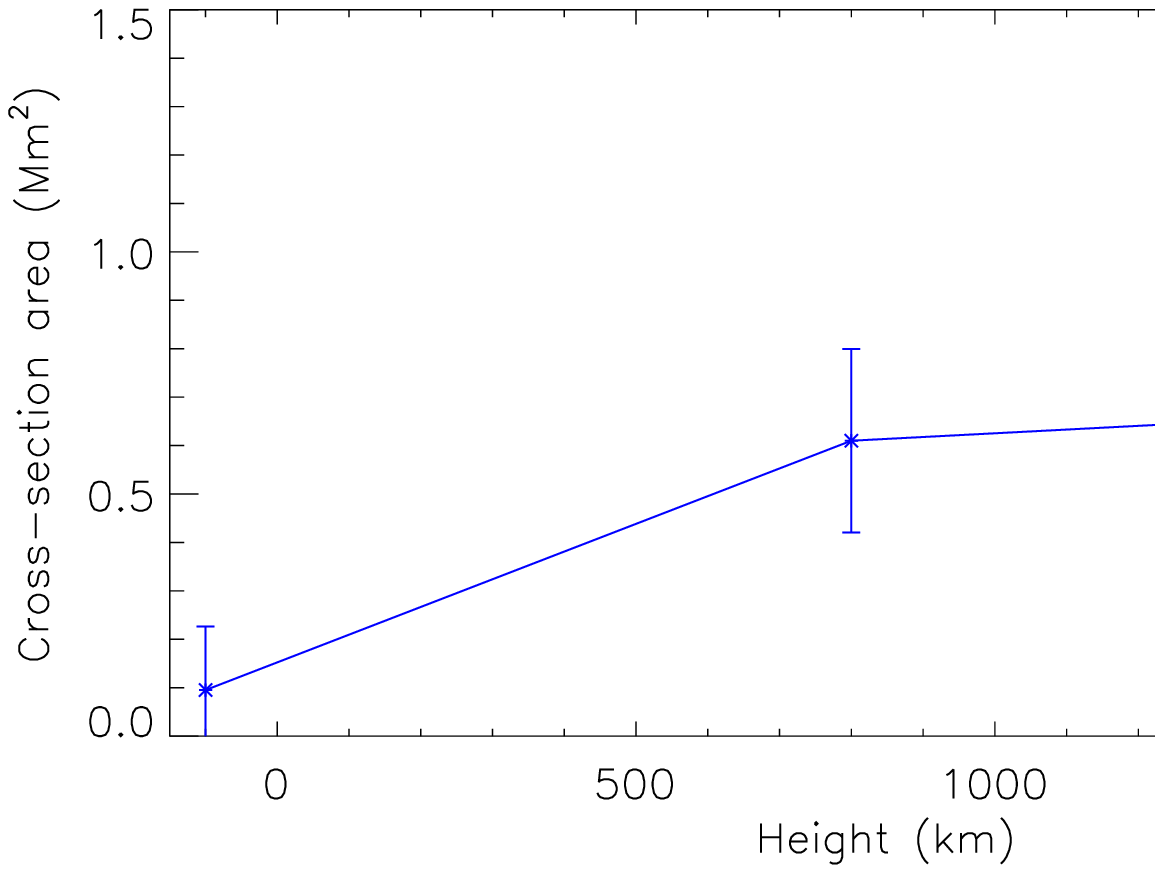} 
    \caption{ Variation of cross-sectional area of loop 6 for 95\% contour level with height. The vertical bars represent errors in the cross-sectional area.} 
    \label{fig:area}
\end{figure}

  Variation of cross-sectional areas of loop 6 at different atmospheric heights (at $\approx$95\% of the maximum correlation values) is shown in Fig.~\ref{fig:area}. Here, we see flux tube expansion with height. Details on the technique and methodology for deriving loop cross-sections are described in \citet{2023MNRAS.525.4815R}.
  Since we are tracing the same loop, we can assume the filling factor to be one from photosphere to corona. However, at the photosphere, loop cross-section is limited to 1-pixel due to limitations on resolution. We have assumed 1- and 2-pixel errors in the width of the loops calculated using SDO and IRIS passbands respectively. 1-pixel of SDO and 2-pixel of IRIS passbands correspond to 435 km and 241 km on the Sun respectively.
 
\subsection{Average umbral Mg II 2796 \AA\ line profile}

\begin{figure}
    \centering
    \includegraphics[width=0.5\textwidth]{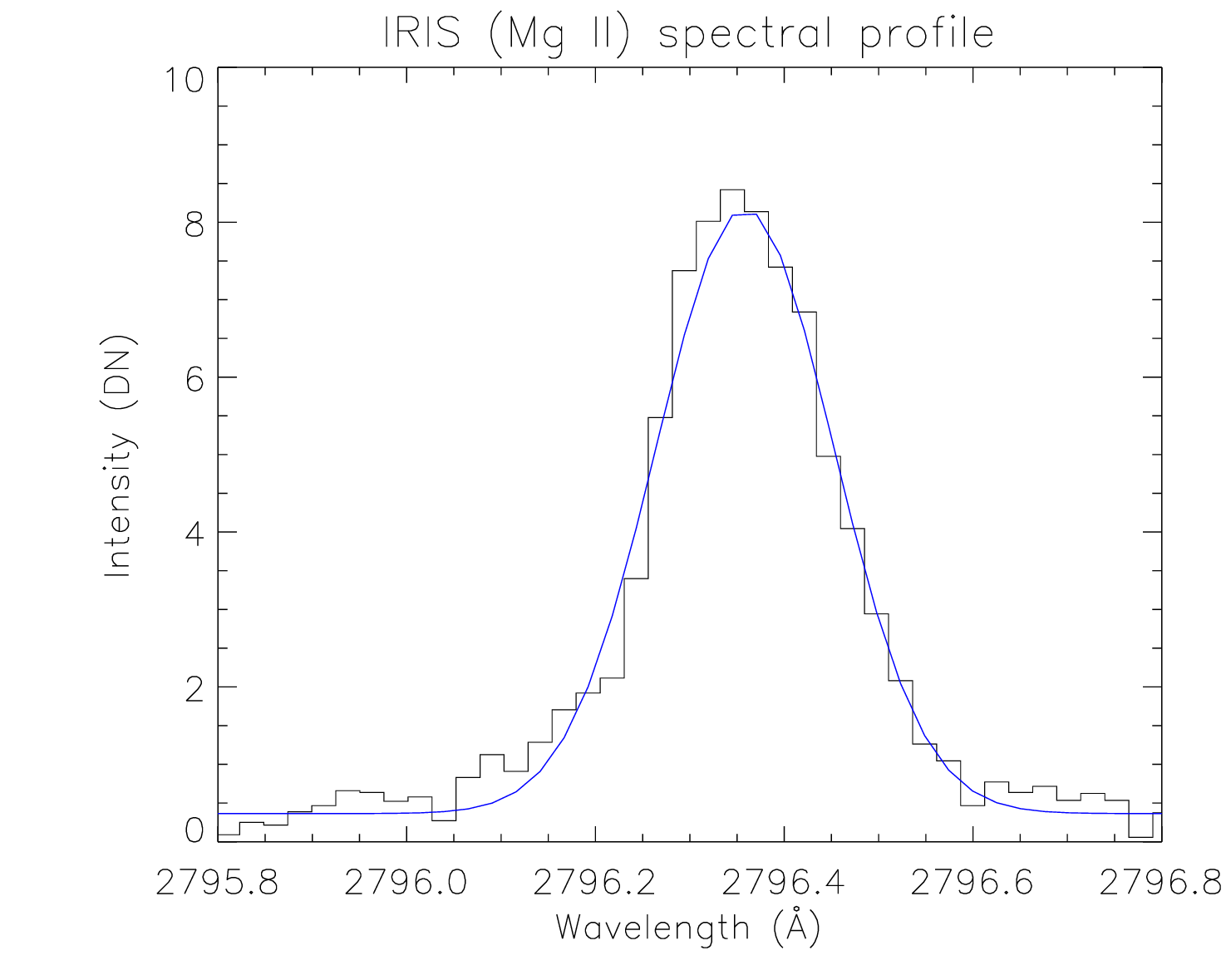} 
    \caption{ Average line profile of Mg II 2796 \AA\ over 11-pixels along the slit and binned over two-time frames at 7:20 UT. The profile is fitted with a Gaussian function in blue color.} 
    \label{fig:mg_line}
\end{figure}

We have extracted the average near-ultraviolet spectra of Mg II 2796 \AA\ in umbral region marked with a white line (11-pixels) in Fig.~\ref{fig:iris}. 
We then fitted average spectral profiles with the Gaussian function and constant background as shown in Fig.~\ref{fig:mg_line}. The fitted parameters were used to extract Doppler shifts and thus Doppler velocity oscillations as shown in the right panel of Fig.~\ref{fig:iris}. 

The fitted parameters were also used to extract peak intensity with time and thus relative intensity amplitude is obtained from spectroscopic data. We then determined the relative intensity amplitude near the slit from imaging data (assuming similar intensity and velocity oscillations) and found that the intensity amplitude obtained from spectroscopic data is 1.25 times larger than the intensity amplitude obtained from imaging data. Therefore, to obtain velocity amplitudes from imaging data ($IV_{amp}$), we are using formula  
\begin{math}
    \frac{SI_{amp}}{SV_{amp}} = \frac{II_{amp} \times 1.25}{IV_{amp}}
\end{math}, where $II_{amp}$ and $SI_{amp}$ are relative intensity amplitude obtained from imaging and spectroscopic data respectively, and $SV_{amp}$ is velocity amplitude obtained from spectroscopic data  for 3-min oscillations.

\label{appendix:spectra}

\subsection{Temperature and density of loop from photosphere to corona}
\label{appendix:temp-dens}

\begin{figure}
    \centering
    \includegraphics[width=0.5\textwidth]{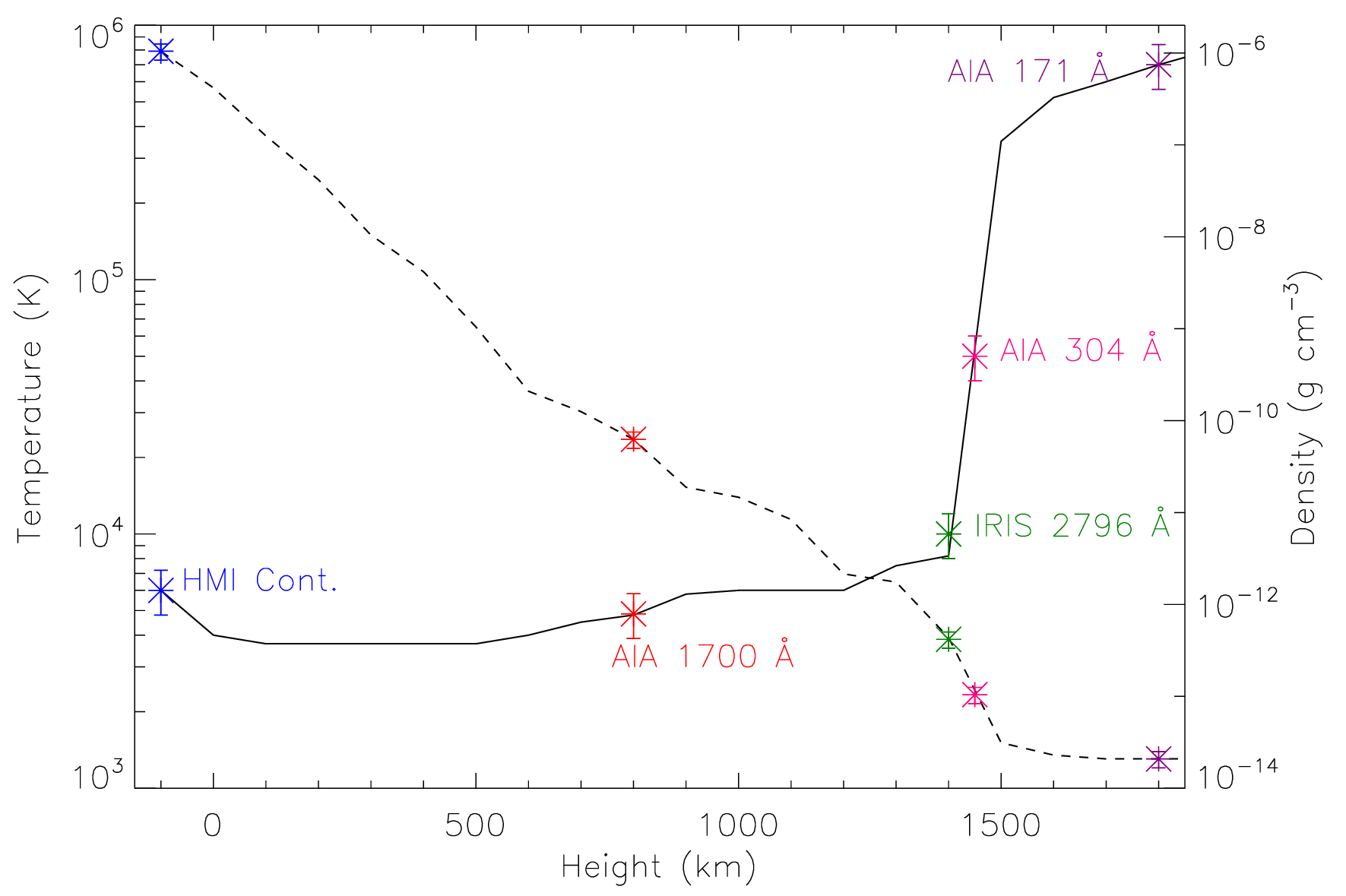} 
    \caption{Temperature and total mass density as a function of height shown by solid and dashed lines respectively for the sunspot model extracted from \citet{1999ApJ...518..480F}. The axis on the left shows the temperature and the right shows the mass density. Different colored asterisk symbols (*) are used to represent different passbands as labeled. }
    \label{fig:fontela}
\end{figure}

 To determine the formation heights for AIA, IRIS and HMI passbands in the sunspot umbra, we utilize the sunspot model of \citet{1999ApJ...518..480F}.  Temperature and density variations with respect to the formation height are shown in Fig.~\ref{fig:fontela}. We associate temperature at the peak of response functions as passbands temperature and obtain the corresponding formation heights. However, all these passbands will also have some cooler and hotter temperature contributions as well \citep[e.g.,][]{2010A&A...521A..21O}. Simultaneously at those formation heights, we determine electron ($N_e$) and hydrogen ($N_h$) number density separately. The mass density is given by $\rho= (N_e+N_h)m_p$, where $m_p$ is the proton mass ($m_p = 1.67\times10^{-24}$ g) \citep{2004psci.book.....A}. The left axis shows the temperature and the right axis shows the total mass density ($\rho$). Furthermore, we have assumed 20\% error in density and temperature in all our calculations \citep[e.g.,][]{2019A&A...627A..62G}. These values play a key role in determining the wave energy fluxes with height. 
 
\subsection{Results from integrated umbra}
\label{appendix:umbra}
\begin{figure*}
    \centering
    \includegraphics[width=0.65\textwidth]{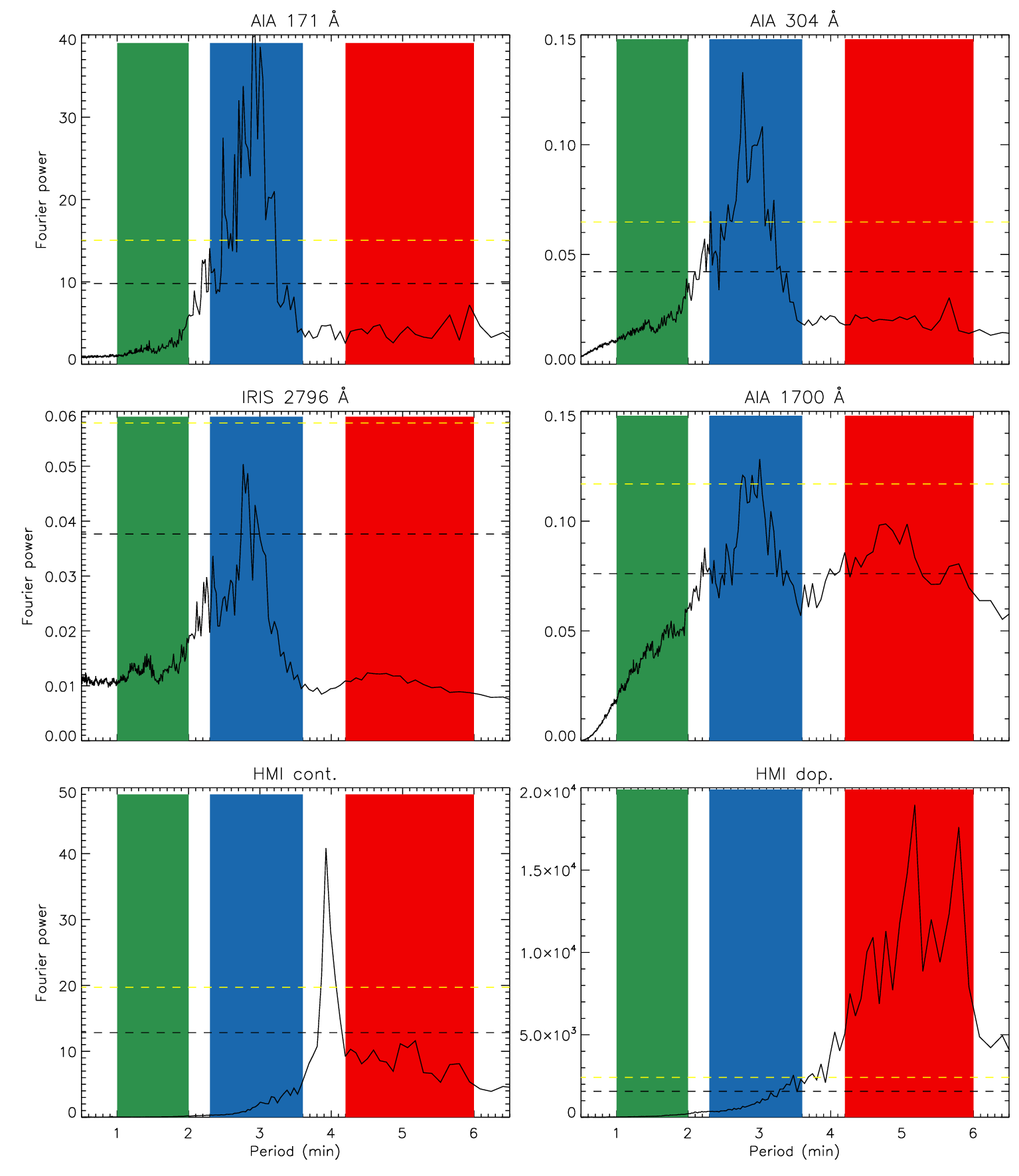} 
    \caption{Average Fourier power spectra obtained from background subtracted light curves at different heights as labeled for integrated umbra shown by contour above HMI continuum image in Fig.~\ref{fig:maps}. Shaded regions in green, blue and red colors denote 1.5-min, 3-min and 5-min period bands respectively identified for our analysis. Horizontal dashed lines in black and yellow colors represent 95\% and 99\% average confidence levels.}
    \label{fig:Power_umbra}
\end{figure*}

\begin{figure*}
    \centering
    \includegraphics[width=0.99\textwidth]{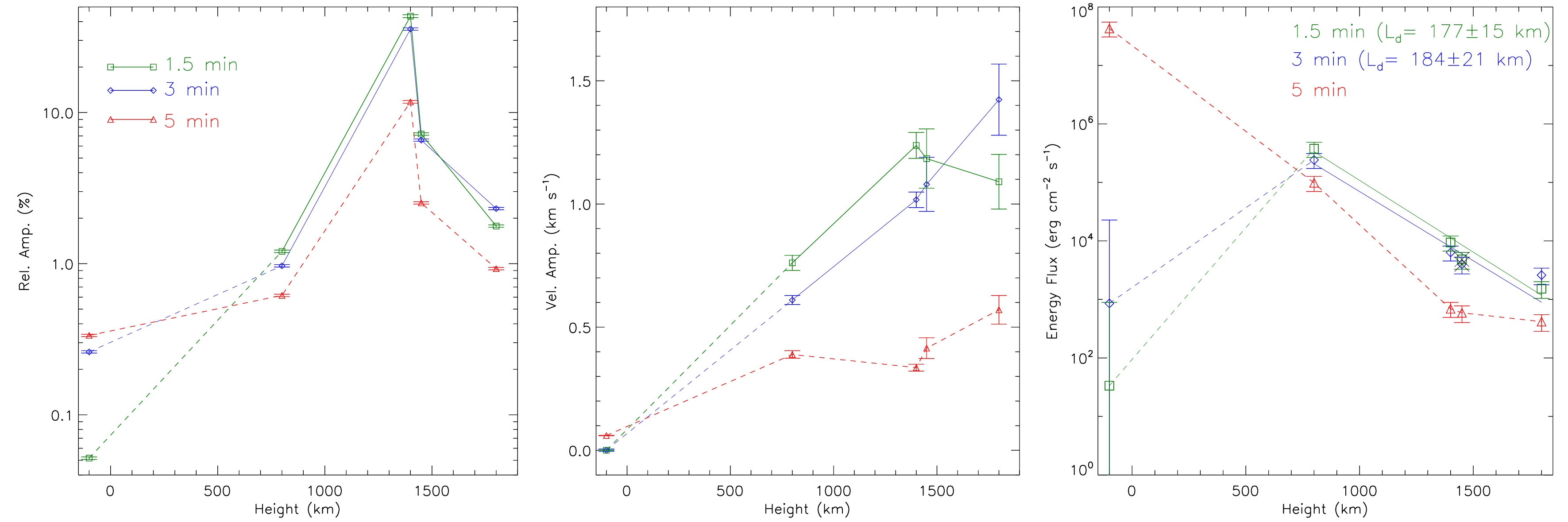} 
    \caption{Variation of average relative intensity amplitudes (left panel), velocity amplitudes (middle panel) and average wave energy fluxes (right panel) with atmospheric heights derived for the whole integrated umbra.}
    \label{fig:umbra}
\end{figure*}

Previously, \citet{2016ApJ...831...24K} and \citet{2017ApJ...847....5K} calculated the wave amplitudes and energy fluxes in the umbral atmosphere over the entire integrated umbra. For comparison with previous results, we also carried out a similar analysis on the integrated umbra (see HMI continuum image of Fig.~\ref{fig:maps}). We determined the average FFT power spectra of the whole umbra at all the atmospheric heights, and obtained the presence of various periods as shown in Fig.~\ref{fig:Power_umbra}. We performed a similar analysis as described earlier to derive relative intensity amplitude, velocity amplitude and wave energy flux in the umbra for three period bands and plot them in Fig.~\ref{fig:umbra}. Due to lack of significant enhancement in the FFT power of 3-min period band at photosphere, we are considering it as an unreliable data point. The damping lengths of wave energy flux for 1.5-min and 3-min period bands obtained from temperature minimum to corona are printed at the top right corner of Fig.~\ref{fig:umbra}. 

\bsp	

\label{lastpage}

\end{document}